\documentclass[graybox]{svmult}

\usepackage{mathptmx}       
\usepackage{helvet}         
\usepackage{courier}        
\usepackage{type1cm}        
\usepackage{makeidx}         
\usepackage{graphicx}        
\usepackage{multicol}        
\usepackage[bottom]{footmisc}
\usepackage{hyperref}        
\usepackage{soul}            
\hypersetup{colorlinks=true,urlcolor=blue}
\usepackage[square,numbers]{natbib}

\graphicspath{{./figs/}}

\usepackage{amssymb}
\usepackage{amsmath}

\usepackage{xcolor}

\makeindex             

\usepackage{journals}
\def\einstein{{\it Einstein}}
\def\rosat{{\it ROSAT}}

\def\chandra{{\it Chandra}}
\def\xmm{{\it XMM-Newton}}
\def\suzaku{{\it Suzaku}}

\def\fermi{{\it Fermi}}

\def\nustar{{\it NuSTAR}}

\def\erosita{{\it eROSITA}}

\def\sgras{{Sgr~A$^\star$}}

\begin{document}
\title*{Diffuse Hot Plasma in the Interstellar Medium and Galactic Outflows}
\author{Manami Sasaki\thanks{corresponding author},
Gabriele Ponti,
and 
Jonathan Mackey}
\institute{Manami Sasaki \at Dr. Karl Remeis Sternwarte, Erlangen Centre for Astroparticle Physics,  Friedrich-Alexander-Universit\"at Erlangen-N\"urnberg, Sternwartstrasse 7, 96049 Bamberg, Germany \email{manami.sasaki@fau.de}
\and Gabriele Ponti \at INAF-Osservatorio Astronomico di Brera, Via E. Bianchi 46, I-23807 Merate (LC), Italy\\ 
Max-Planck-Institut f{\"u}r extraterrestrische Physik, Giessenbachstrasse, D-85748, Garching, Germany \email{gabriele.ponti@inaf.it}
\and Jonathan Mackey \at Dublin Institute for Advanced Studies, Astronomy \& Astrophysics Section, Dunsink Observatory, Dunsink Lane, Dublin D15 XR2R, Ireland. \email{jmackey@cp.dias.ie}
}
%
%
\maketitle
\abstract{
We summarise observations and our current understanding of the interstellar medium (ISM) in galaxies, which mainly consists of three phases: cold atomic or molecular gas and clouds, warm neutral or ionised gas, and hot ionised gas. These three gas phases form thermally stable states, while disturbances are caused by gravitation and stellar feedback in form of photons and shocks in stellar winds and supernovae. Hot plasma is mainly found in stellar bubbles, superbubbles, and Galactic outflows/fountains and is often dynamically unstable and is over-pressurised. In addition, in galactic nuclear regions, accretion onto the supermassive black hole causes enhanced star formation, outflows, additional heating, and acceleration of cosmic rays. 
}

\vspace{\baselineskip}

\section{The hot phase of the ISM}

The interstellar medium (ISM) near the midplane of the Galaxy is characterised by at least three different phases \citep{McKOst77}.
The cold atomic/molecular phase occupies a small fraction of the volume but has significant mass, with typical gas number-density ranging from $n \sim 10^2-10^5$\,cm$^{-3}$, and gas temperature $T$, ranging from $T\sim10^2-10$\,K.
The warm neutral/ionized medium occupies a significant fraction of the Galactic midplane both in terms of mass and volume, with density $n \sim 10^{-1}-10$\,cm$^{-3}$ and temperature $T\sim10^4-10^2$\,K.

The hot ISM has low density, with most of the hot volume having $n\lesssim10^{-2}$\,cm$^{-3}$ and $T\gtrsim10^6$\,K.
It occupies a large (but uncertain) fraction of the volume of the Galactic midplane and, while the topology of the occupied volume is also quite uncertain, it certainly consists of bubbles, superbubbles and Galactic fountains \citep{DeABre04, Cox05, HopQuaMur12, RatNaaGir21}.
The \emph{Local ISM} around the Sun provides some of the best observational data, and we now know that the Sun is moving through a medium with warm, diffuse clouds embedded in a warm/hot \emph{Local Bubble}, with a mean thermal pressure $P/k\approx3\,000-10\,000$\,cm$^{-3}$\,K \citep{FriRedSla11} (where $P$ is the gas pressure and $k$ is the Boltzmann constant).
Diffuse thermal X-rays with $kT\approx0.1$\,keV are measured from the Local Bubble \citep{LiuChiCol17} and provide a foreground that must be accounted for in studies of the hot ISM at larger distances from the Sun.

In the vicinity of active energy sources such as star clusters, black holes or recent supernovae, the hot ISM can be strongly overpressurised with respect to the background ISM.
For example the diffuse hot gas within the young and massive star cluster Westerlund 1 has $n_\mathrm{p}\sim1$\,cm$^{-3}$ and $T\sim10^7$\,K \citep{KavNorMeu11}, i.e., $P/k\sim10^7$\,cm$^{-3}$\,K, about $1000\times$ larger than the typical thermal pressure in the diffuse ISM of the Galactic midplane.

Closer to the Sun, a recent study \citep{ZucGooAlv22} shows that all star forming clouds and young stellar associations within 200\,pc of the Sun are located at the outer edge of the Local Bubble and expanding with the bubble.
Their work confirms long-standing ideas, showing that the Local ISM properties are largely determined by an episode of massive-star formation (followed by supernova explosions) that began about 14\,Myr ago and energised the local warm/hot ISM.
The overpressurised bubble expanded for millions of years until it reached pressure equilibrium with the surrounding Galactic Plane, but still young stars and gas clouds are expanding with the momentum that was imparted to them through this expansion.

\begin{figure}
    \centering
    \includegraphics[width=1.0\textwidth]{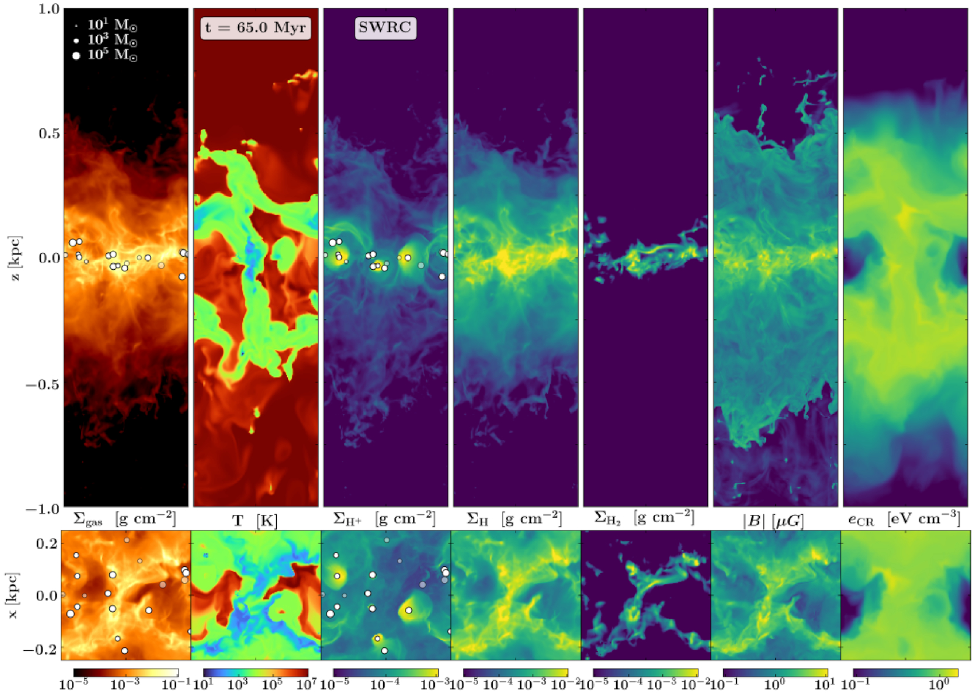}
    \caption{The multi-phase ISM from 3D simulations of the Galactic disk by \citet{RatNaaGir21}.  Reproduction of Fig.~1, ``SILCC VI - Multiphase ISM structure, stellar clustering, and outflows with supernovae, stellar winds, ionizing radiation, and cosmic rays,`` (Rathjen \emph{et al.}, 2021, MNRAS, 504, 1039, \copyright{} the authors).
    \label{fig:multiphase}}
\end{figure}

The three ISM phases, cold, warm and hot, are located in thermally stable regions of parameter space in the pressure-density plane \citep{McKOst77}, and so gas does not easily move from one phase to the other.
Thermal stability arguments initially led many researchers to consider an equilibrium model of the Galactic plane, but this view was not supported by numerical simulations \citep{DeABre05,WalGirNaa15, KimOst17} or studies showing rapid formation and short lifetimes of molecular clouds \citep{HarBalBer01, IbaMacKle16, HopQuaMur12}.
The modern view of the Galactic ISM is of a dynamic medium, constantly stirred and disturbed by the interaction between gravity and stellar feedback in the form of winds, photo-heating, supernovae and cosmic rays \citep{RatNaaGir21} (see Fig.~\ref{fig:multiphase}).
On sufficiently large scales the system can be characterised as a dynamic equilibrium, where statistical properties of the medium are unchanging.
The fraction of the mass in thermal equilibrium is quite large because the cold and dense gas has a short cooling timescale, but a significant fraction of the ISM volume is not in thermal or ionization equilibrium \citep{DeABre05}.

Heating and ionization by far- and extreme-UV (FUV, EUV) radiation fields can effectively heat gas from the cold to the warm phase, but generally strong shocks are required to heat the cold/warm gas and add it to the hot phase.
The post-shock temperature of a strong adiabatic shock of velocity $v_\mathrm{sh}$ is 
\begin{equation}
T_\mathrm{ps} = \frac{3\mu}{16k}v_\mathrm{sh}^2 = 1.4\times10^7\,\mathrm{K}\, \left( \frac{v_\mathrm{sh}}{1000\,\mathrm{km\,s}^{-1}} \right)^2  \;,  \label{eqn:tps}
\end{equation}
where $\mu\approx1.0\times10^{-24}$\,g is the mean mass per particle in an ionized plasma with typical ISM abundances of H and He, and we have assumed an adiabatic index $\gamma=5/3$.

Slow shocks ($v_\mathrm{sh}\lesssim1000$\,km\,s$^{-1}$) in the cold or warm ISM are strongly radiative and cannot add to the hot phase.
Above $T_\mathrm{ps}\approx3\times10^7$\,K ($\approx3$\,keV, $v_\mathrm{sh}\gtrsim1500$\,km\,s$^{-1}$), however, the only effective radiative cooling mechanism for the thermal plasma is bremsstrahlung \citep{SutDop93}, for which the cooling rate has the scaling $\dot{E}_\mathrm{ff}\propto n_e^2 \sqrt{T}$, where $n_e$ is the electron number density.
The cooling time is $t_c = E/\dot{E}_\mathrm{ff} \propto n_e^{-1} \sqrt{T}$, increasing with temperature.
Typically $t_c \sim10^6-10^9$ yr for gas with $T\gtrsim3\times10^7$\,K \citep{SutDop93}, longer than the lifetime of supernova remnants and clusters of massive stars.

The hot phase of the ISM is therefore effectively adiabatic except for energy loss through the boundary layer that interfaces with the much denser cooler phases.
Thermal conduction \citep{CowMcK77} acts to broaden the boundary layer between hot and cold gas, increasing the volume of gas with intermediate temperatures ($10^5-10^7$\,K) where radiative cooling is most effective.
Similarly, turbulent mixing can have the same effect, driven by shear flows \citep{SlaShuBeg93}. 
This was already recognised as important to the overall structure of the Galactic ISM by \citet{McKOst77}, although in more recent theoretical models the steady-state conductive interface is replaced by a dynamically mixed layer \citep{NakMcKKle06, MacGvaMoh15, LanOstKim21}.

\section{Sources of the hot ISM}

The main sources of shocks with $v_\mathrm{sh}\gtrsim1500$\,km\,s$^{-1}$ in the Galaxy are supernovae and winds from hot stars.
Near the Galactic Centre, episodic accretion onto the supermassive black hole, Sgr A*, might also drive energetic outflows that shock-heat their surroundings, generating hot plasma and enhanced cosmic ray density.  Sgr A* was also the first confirmed Galactic PeVatron \citep{2016Natur.531..476H}.
On smaller scales, protostellar jets, novae and winds from low-mass stars can contribute to creating gas with $T>10^6$\,K and accelerating particles. 
Extreme objects such as pulsar wind nebulae, high-mass X-ray binaries and microquasars like SS433 also contribute to the hot ISM, and may contribute very significantly to the highest energy cosmic rays in the Galaxy, 
but the total energy input of these objects to the Galaxy is low compared with supernovae and stellar winds
(see chapter on Galactic cosmic rays by Strong, Merten, Evoli)


Simulations and observations show that large-scale superbubbles and Galactic fountains are produced by the combined effects of winds and supernovae (and associated CR production) from many stars in young star clusters.
These are the main sources of the hot, diffuse ISM, and the main drivers of its dynamics.
While they are good targets for X-ray observations on account of their large emission measure, their complexity makes interpretation of observational data difficult.
Moreover, what we see today is the integrated history of energy input from stellar winds and supernovae over millions of years, from uncertain initial conditions.
Often it is the smaller and simpler structures, with shorter dynamical timescales, such as isolated supernova remnants, colliding-wind binary systems, novae, stellar wind bubbles and bow shocks, where the important physical processes can be investigated in detail.
In these systems a quantitative comparison between theory and observation is possible and can give strong constraints that do not depend on the uncertain history of the source.

\subsection{Stellar wind bubbles and bow shocks}

\begin{figure}
    \centering
    \includegraphics[width=0.8\textwidth]{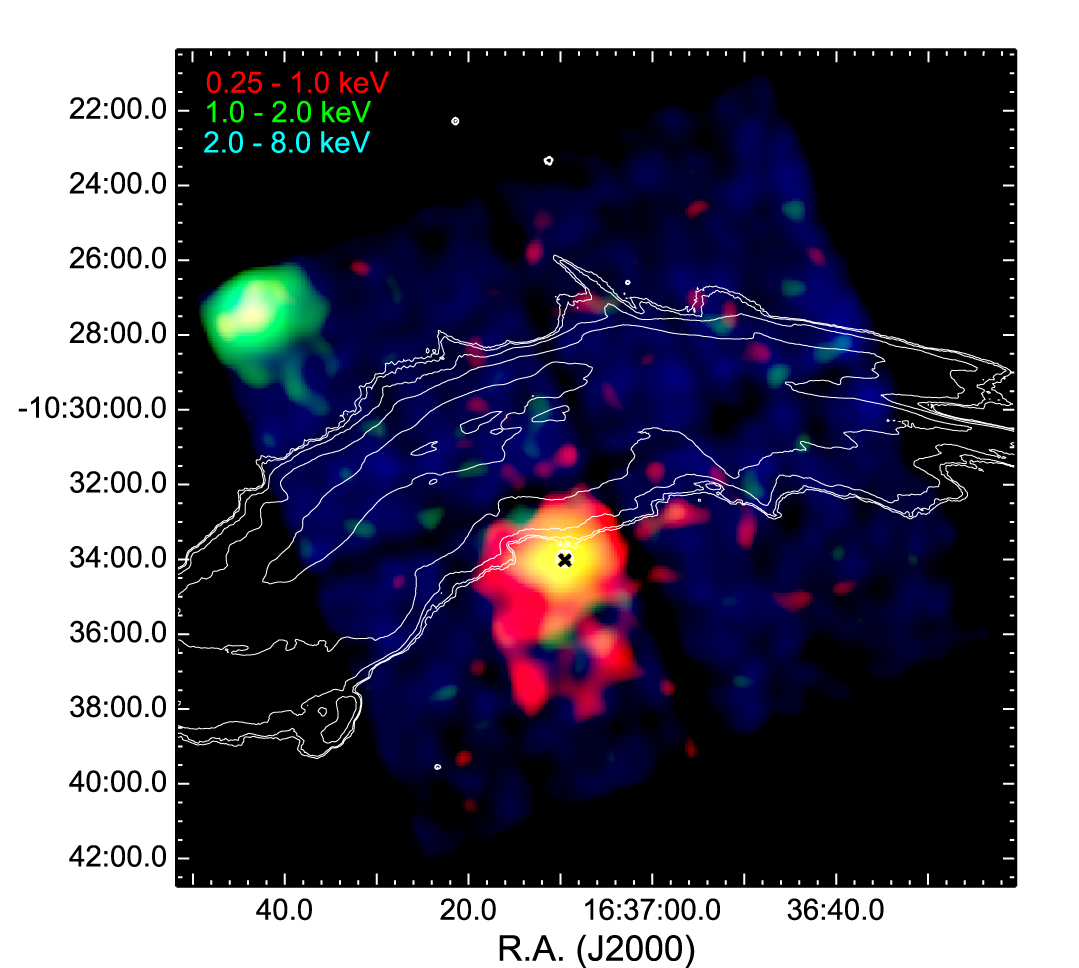}
    \caption{Diffuse X-ray emission from the shocked stellar wind around the O star $\zeta$ Oph, including a contour plot of the infrared bow shock.  This is a reproduction of figure 3 (4th panel) from \citep{ToaOskGon16} (\copyright{} AAS. Reproduced with permission. \href{https://doi.org/10.3847/0004-637X/821/2/79}{https://doi.org/10.3847/0004-637X/821/2/79}).
    \label{fig:zetaoph}}
\end{figure}

Runaway massive stars, dynamically ejected from their place of birth, are typically moving through the diffuse ISM and their wind bubbles drive bow shocks when the star is moving supersonically.
Among these are massive O-type and Wolf-Rayet (WR) stars, with some of the most extreme stellar winds of any stars in the Galaxy, and it was recognised very early on that these should be surrounded by X-ray emitting wind bubbles \citep{1975ApJ...200L.107C}.

Observations with \emph{Chandra} and \emph{XMM-Newton} have detected thermal X-ray emission from 3 bubbles around WR stars: S\,308, NGC\,2359 and NGC\,6888 \citep{ToaGueChu16}.
The plasma temperature of $\approx10^6$\,K in all cases is well below that predicted by Equation~\ref{eqn:tps}, implying significant mixing of denser gas into the hot bubble through either thermal conduction or turbulent mixing \citep{ToaArt18}.
Uncertainty in the previous evolutionary history of the WR stars complicates the interpretation of these measurements \citep{ToaGueChu16}, but clearly there are significant energy losses from the hot phase through the interface with cooler and denser surrounding gas.
This agrees with results from surveys of wind-blown bubbles around isolated stars and star clusters \citep{RosLopKru14, OliLopRos21}.

For isolated O stars, diffuse thermal X-ray emission was found around the nearest O star, $\zeta$ Oph \citep{ToaOskGon16}, and also within the H~\textsc{ii} region RCW\,120 \citep{TowBroGar18}.
Predictions that the Bubble Nebula, NGC\,7635, could emit detectable diffuse X-ray emission \citep{GreMacHaw19} proved over-optimistic \citep{ToaGueTod20}, reinforcing again that energy losses from wind bubbles efficiently depressurize the X-ray emitting plasma.
Bow shock nebulae such as NGC\,7635 or the nebulae around $\zeta$ Oph or BD+43$^\circ$3654 \citep{2012PASJ...64..138T} are nevertheless useful targets for further investigation because of their geometric simplicity.
The time independence of the bow shock (to first approximation) also removes the requirement to solve an inverse problem when interpreting observations, differently from superbubbles and supernova remnants.


Non-thermal emission from wind bubbles around isolated massive stars has proven elusive, with only upper limits so far published at X-ray or gamma-ray energies, and radio synchrotron detections only from the bow shocks of BD+43\,3654 \citep{BenRomMar10} and BD+60\,2522 \citep{MouMacCar22}, as well as the nebula around the WO star WR\,102 \citep{PraTejDel19}.
There is potential for more radio detections, but measurements of high-energy emission will likely require the next generation of observatories such as the Cherenkov Telescope Array (CTA) \citep{2019scta.book.....C}.

Wind-wind collisions in binary systems offer another view of the contribution of stellar winds to the hot ISM, in both thermal and non-thermal emission.
The shock speeds are essentially the same as termination shocks in isolated wind bubbles, but the shocks are orders-of-magnitude denser (and hence brighter) because they occur much closer to the driving stars.
Non-thermal radio emission from the colliding wind binary system WR140 was explained in the context of particle acceleration in the wind collision shocks \citep{1993ApJ...402..271E}, and it was suggested that gamma-ray emission should also be detectable.
Since then synchrotron radio emission has been detected from many binary systems containing massive stars, especially Wolf-Rayet (WR) stars \citep{DouWil00}.
Gamma-ray emission has also been detected from the closest WR binary system to Earth, $\gamma^2$ Velorum \citep{Psh16, MarRei20} and from the most extreme Galactic binary system known, $\eta$ Carinae \citep{TavSabPia09, HESS20_EtaCar}.
These results show that stellar-wind shocks are efficient particle accelerators to very high energies and suggest that there are many more gamma-ray sources among the colliding-wind binaries that will be detected by more sensitive observations with e.g.\ CTA.
At the same time, radio observations at low frequencies allow to detect more wind-driven bow shocks \citep{2022MNRAS.510..515V}, and next generation facilities will see a big increase in sensitivity to wind bubbles and binary systems.
The coming decade should see a dramatic improvement in our understanding of the shock physics in stellar-wind bubbles.

\subsection{Supernova remnants}

Soon after the first observations in the X-ray band carried out on rocket flights revealed 
cosmic X-ray sources other than Sun 
\cite{1962PhRvL...9..439G},
it has been shown that one of the major sources of cosmic X-rays 
is low-density plasma at temperatures of the order of 10$^7$ K
like in supernova remnants (SNRs)
\citep[see, e.g.,][and references therein]{1964ApJ...140..470H,1965PhRvL..14..771M,
1976ARA&A..14..373G}
(see also chapters by Bamba and Williams; Yamaguchi; Bamba and Vink).

To understand the properties of the interstellar plasma and its
relation to the colder phases of the ISM, it is important to study the entire SNR population in a galaxy. 
The soft X-ray emission from thermal interstellar plasma, however, is easily absorbed by interstellar
matter on the line of sight. Therefore, it is difficult to study the hot phase of the 
ISM in our Galaxy. 
The two largest satellite galaxies of the Milky Way, Large and Small Magellanic Cloud
(LMC, SMC), with their low foreground absorption and small distances of
$\sim$50~kpc and 60~kpc, respectively, 
are the ideal laboratories for exploring the global structures of the ISM in a galaxy. 
Studies of different types of SNRs and a systematic study 
of the SNR populations in the LMC and SMC have been performed in radio, optical, and X-rays
\citep[e.g.,][]{2016A&A...585A.162M,2017ApJS..230....2B,2021MNRAS.500.2336Y},
yielding $\sim$60 confirmed SNRs in the LMC and many more candidates.
With stellar masses of $2.7 \times 10^{9} M_\odot$ \citep{2006lgal.symp...47V} and $6 \times 10^{10} M_\odot$ \citep{2015ApJ...806...96L} 
in the LMC and the Milky Way, respectively, one would expect about one order of magnitude higher number of SNRs in the Milky Way. Compared to $\sim$300 known SNRs and candidates in the Milky Way \citep{2019JApA...40...36G}, the sample in the LMC is more complete.
Many of the SNRs were found to be evolved SNRs in the Sedov phase of evolution, in which they are expanding adiabatically. By comparing the observational data with the Sedov dynamical model, physical properties of the SNRs were determined including expansion velocity, age, and explosion energy. Furthermore, a new class of evolved SNRs has been discovered in the Magellanic Clouds with an X-ray bright, Fe-rich core, consistent with reverse shock-heated Type~Ia ejecta.
A significant incompleteness in the X-ray luminosity function below $\sim$8$\times10^{34}$~erg~s$^{-1}$ was found for LMC SNRs \citep{2016A&A...585A.162M}.
So far, only 3 SNRs have been detected with a luminosity below 10$^{34}$~erg/s in the LMC and one in the SMC, 
while both Magellanic Clouds are the only galaxies in which we can probe the X-ray emission of SNRs down to the lowest luminosities.







\subsection{HII regions and superbubbles}

Massive stars emit strong ultraviolet photons due to their high temperature and 
form HII regions around them, in which the hydrogen gas is photoionised. 
In addition, optical and ultraviolet observations of massive stars have shown that they 
emit strong stellar winds
\citep{1967ApJ...150..535M,1970ApJ...160..595S,1979ARA&A..17..275C}.
These winds will cause high-velocity shock waves, which expand into the circumstellar medium and 
create a thin dense circumstellar shell around a stellar bubble
\citep[see][and references therein]{1975ApJ...200L.107C}.
The combination of stellar winds of many massive stars and their supernova explosions, which heat and ionise the ambient medium, creates
larger structures in the ISM. These objects are called
superbubbles and are filled with
low-density,
high-temperature interstellar plasma.
In particular, before any observation of such a high-temperature plasma was possible, it was suggested that the bright and nearby HII region Rosette Nebula 
was produced by the action of stellar winds of massive stars in the central cavity
\citep{1966ApJ...144..206M}. 

When first X-ray observations were carried out, diffuse X-ray emission was detected around  the Trapezium stars in the 
the Orion nebula in the energy band of 0.1 -- 4 keV with the imaging proportional counter (IPC) aboard the \einstein\ Observatory  and 2 -- 11 keV with the  rotation modulation collimators on SAS-3
(\citep{1979ApJ...234L..59K,1979ApJ...228L..33B}, respectively).
Observations with the \einstein\ X-ray Observatory and \rosat\ have also revealed soft diffuse X-ray emission inside HII regions and around OB associations in the LMC (e.g., \cite{1990ApJ...365..510C,1991ApJ...373..497W}). These first studies have shown that the X-ray emission detected from superbubbles are brighter than what is predicted by the standard model for stellar wind bubbles suggested by \cite{1975ApJ...200L.107C,1977ApJ...218..377W}.
Different scenarios have been suggested which will result in increasing the flux, such as heating by shocks of SNRs inside a superbubble \cite{1990ApJ...365..510C,1995ApJ...450..157C} or enhanced element abundances due to supernovae \cite{2001MNRAS.324..191S}.
Only the X-ray faint superbubbles have most likely not experienced additional input of energy, and are thus ideal targets to test bubble models \cite{1995ApJ...450..157C}. 
In addition, it was shown that X-ray bright superbubbles have faster expansion velocities, while X-ray faint superbubbles have dynamics consistent with what is expected from their stellar populations \cite{1996ApJ...467..666O}.
Many superbubbles also show evidence of break-out with X-ray emission extending beyond the H II shell \cite{2001ApJS..136..119D}.
In addition, it was shown that HII regions in the LMC form supergiant
shells on the inside edges of HI shells. Inside the cavities surrounded by the HI shells, 
there are populations of  massive O and B stars, which have most likely created the supergiant 
shells by their stellar winds and supernovae
\citep{1978A&A....68..189G}.

\begin{figure}[t]
    \centering
    \includegraphics[width=0.32\textwidth]{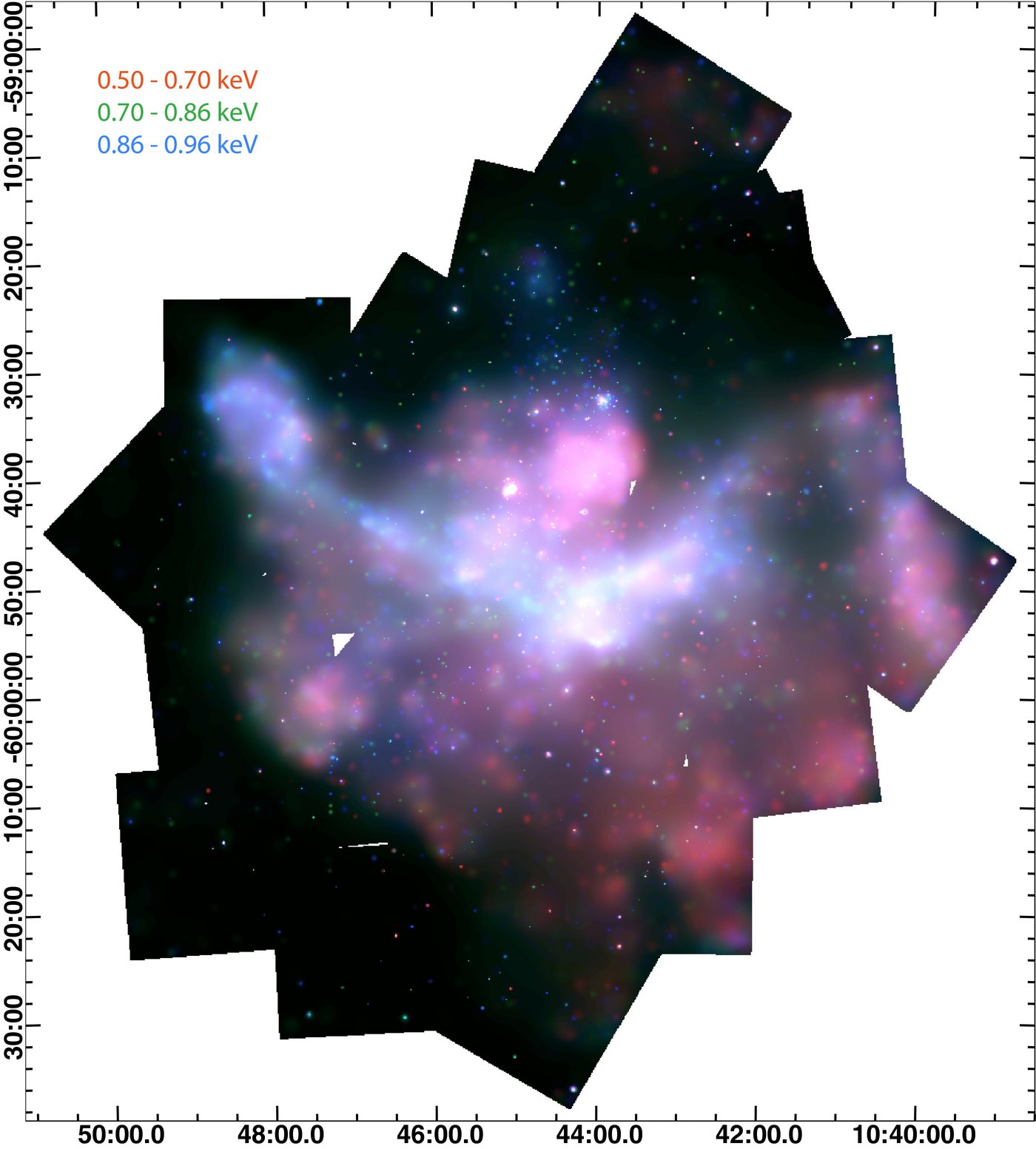}
    \includegraphics[width=0.67\textwidth]{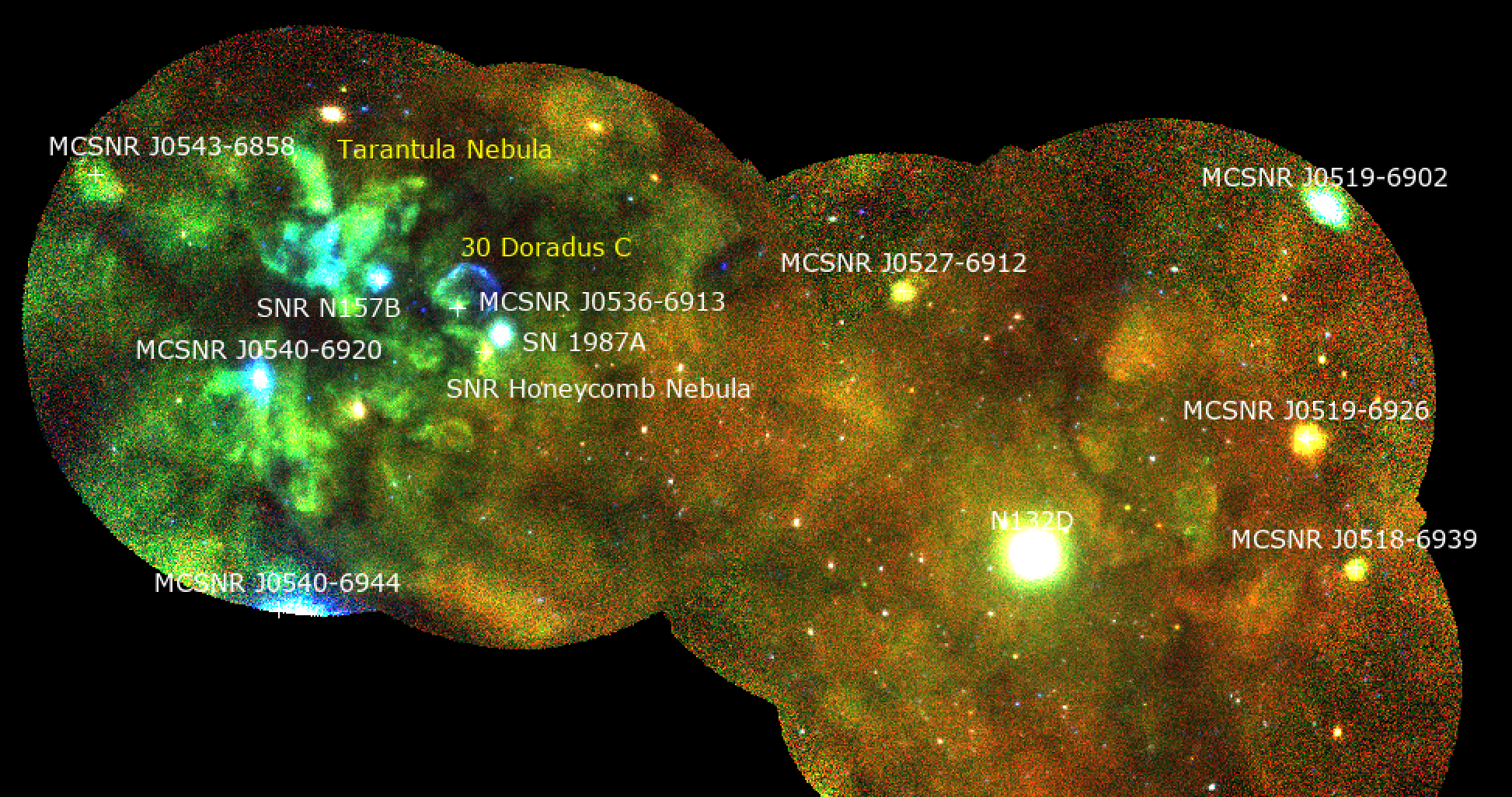}
    \caption{Left: \chandra\  mosaic of the \chandra\  Carina survey \cite{2011ApJS..194....1T} (Townsley et al., ApJS, 194, 1, 2011, reproduced by permission of the AAS).
    Right: Mosaic of \erosita\ images taken during the performance verification/calibration phase (red: 0.2–0.5 keV, green = 0.5–1.0 keV, blue = 1.0–2.0 keV). SNRs are marked in white. Adapted from \citet{2022A&A...661A..37S}.
    \label{fig:xrayimages}}
\end{figure}

Newer X-ray telescopes with better spatial and spectral resolving power have made detailed
X-ray studies of superbubbles possible
\citep[e.g.,][]{1990ApJ...365..510C,1991ApJ...374..475W}.
In particular, since the launch of \xmm\  and \chandra\  X-ray telescopes equipped with X-ray charge-coupled devices (CCDs) in 
1999, which are ideal telescopes for studying extended, soft X-ray 
sources thanks to their high sensitivity and improved spatial and spectral resolution, we are able to carry out studies of the properties of the hot interstellar 
plasma and its origin.
A complete survey of the Carina Nebula was carried out with \chandra\ 
\citep{2011ApJS..194....1T} (see Fig.\,\ref{fig:xrayimages}, left). It led to the detection and studies of X-rays from a large number of young massive stars and star clusters as well as diffuse emission in the entire nebula.
In the LMC, the 30 Doradus region was studied in detail using \chandra\  \citep{2006AJ....131.2140T}, which allowed to resolve the complex structures in the giant HII region.
Other superbubbles and the hot phase in the ISM in the Magellanic Clouds as well as in the Andromeda galaxy (M31)
have been studied with \xmm\ 
\citep[e.g.,][]{2011A&A...528A.136S,2012A&A...547A..19K,2020A&A...637A..12K}.

In these studies, the properties of the hot interstellar plasma were derived from spectral analysis, the energy budget of the superbubbles were compared to the input by the underlying stellar populations, and additional supernova remnants were found.
The diffuse X-ray emission from the ISM of normal galaxies requires at least two thermal plasma components: a lower temperature of $kT \approx$ 0.2 keV and a hotter component with $kT >$0.5 keV
\citep[e.g.][and references therein]{2010ApJS..188...46K,2020A&A...637A..12K}.
The lower-temperature plasma is most likely the hot phase of the ISM in equilibrium, while the additional hotter component is found in regions that have been heated recently (i.e. superbubbles and SNRs) or include unresolved sources like X-ray emitting binaries.  
The density and pressure of the hot interstellar plasma derived from the fit of the lower-temperature spectral component are consistent with those in the Milky Way.
First study of the hot phase of the ISM in the LMC with \erosita, which has similar
resolution to \xmm\  but a much larger field of view, has been
presented by 
\citet{2022A&A...661A..37S} (see Fig.\,\ref{fig:xrayimages}, right). The diffuse emission is well reproduced with two thermal plasma components. The flux of the diffuse emission is higher in the 30 Doradus region compared to the surrounding ISM, indicative of heating by the young massive stars. However, the emission appears darker (and harder, see Fig.\,\ref{fig:xrayimages}, right) since the absorbing column density is higher, as confirmed by the spectral analysis.

\begin{figure}[t]
    \centering
    \includegraphics[height=.22\textheight]{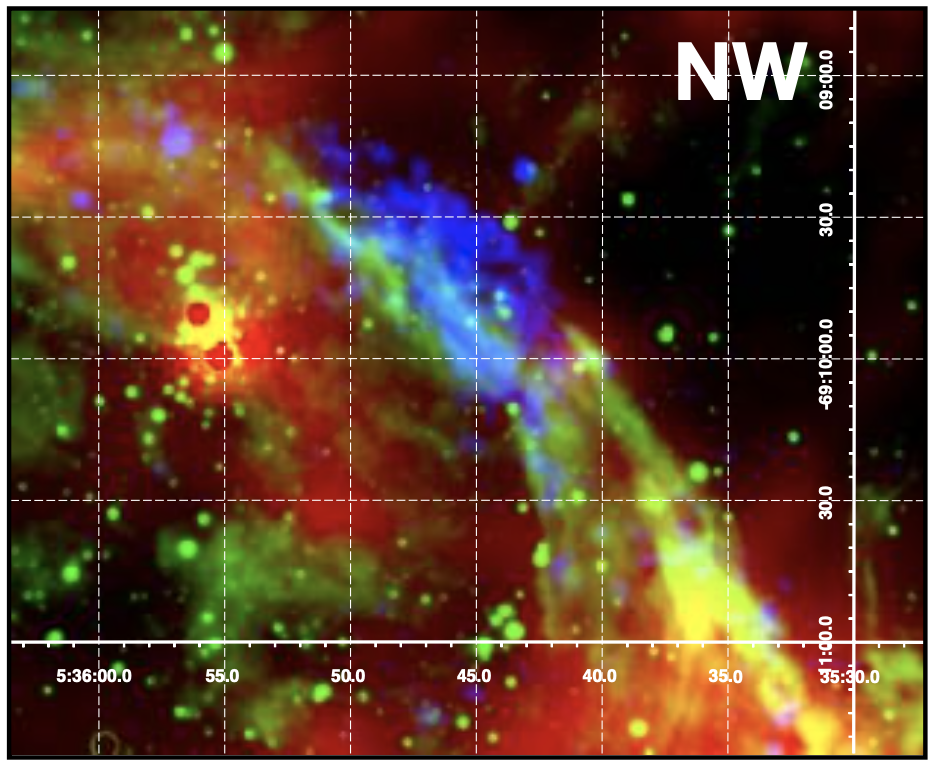}
    \includegraphics[height=.22\textheight]{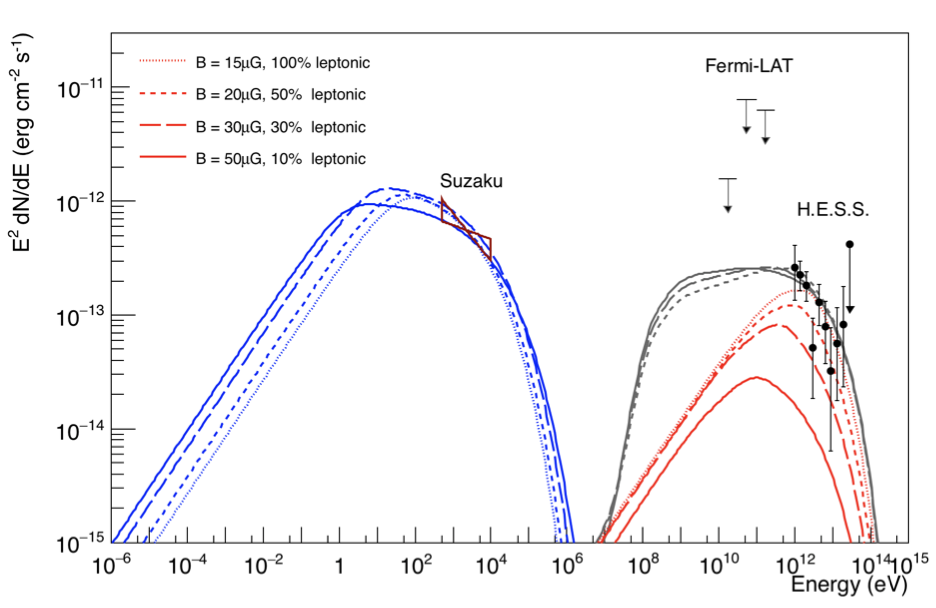}
    \caption{Left: Northwestern shell of 30 Dor C in three colours with red = 24 $\mu$m, green = H$\alpha$, blue = X-rays (1.5--8 keV).
    Right: Spectral energy distribution of 30 Dor C showing constraints about the non-thermal emission in X-ray derived from \suzaku, GeV emission from \fermi-LAT, and TeV emission from H.E.S.S. \cite{2015Sci...347..406H}. The blue lines show the models for the synchrotron component, the red lines show the contribution of inverse-Compton emission, and the black lines show the combination of inverse-Compton and hadronic components.
    Credit: Kavanagh et al., A\&A, 621, 138, 2019, reproduced with permission \textcopyright ESO.
    \label{fig:30dorc}}
\end{figure}

While the emission from superbubbles is typically caused by the hot thermal plasma in their interiors, there is one  superbubble in the LMC, 30 Dor C, with an X-ray spectrum that is dominated by non-thermal synchrotron emission
\citep{2004ApJ...602..257B}. 
\nustar\ observations have shown that the complete shell of 30 Dor C is detected up to 20 keV \citep{2020ApJ...893..144L}.
30 Dor C is the only  non-thermal superbubble known in the Milky Way system 
(with IC 131 in M33 being the only other superbubble confirmed so far to emit non-thermal X-rays \citep{2009ApJ...707.1361T}) and is also a TeV source.
Measurements of the X-ray flux and the magnetic field strength derived from the extent of the non-thermal filaments of the outer shells of 30 Dor C observed with \xmm\  and \chandra\  suggest that the TeV emission of 30 Dor C is caused by inverse Compton scattering of photons on highly relativistic electron, which have been accelerated in strong shocks 
in the superbubble
\citep{2015A&A...573A..73K,2019A&A...621A.138K} (see Fig.\,\ref{fig:30dorc}).
The shocks are likely formed in winds of the young massive stars, in turbulence caused by the combination of 
the winds, and possibly in recent supernovae that occurred inside the superbubble.

\begin{figure}[t]
    \centering
    \includegraphics[width=\textwidth]{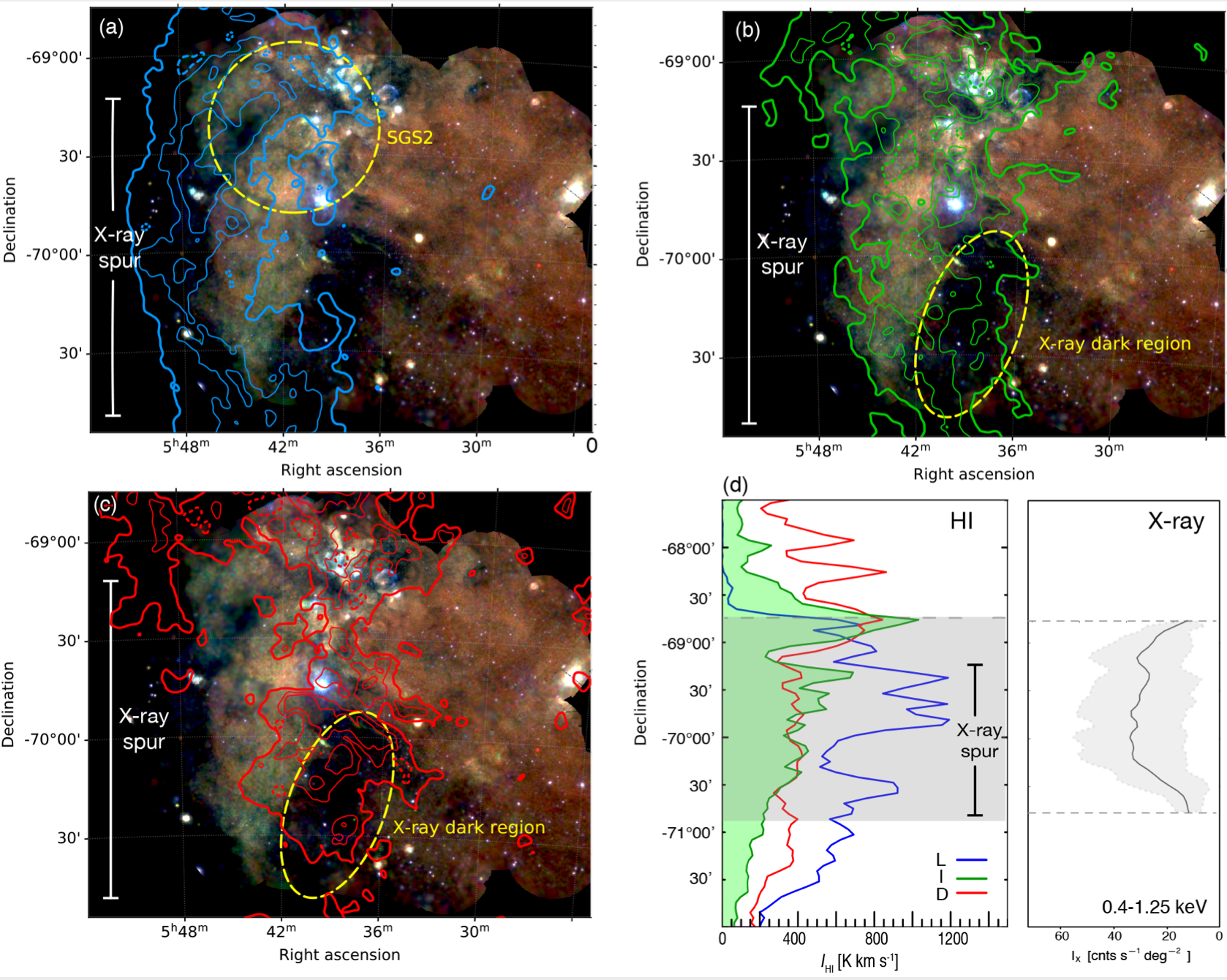}
    \caption{\xmm\ three-color mosaic of the X-ray spur with H I contours for the L-component (a), I-component (b), and D-component (c). The X-ray colors correspond to red: 0.4 -- 0.7 keV, green: 0.7 -- 1.0 keV, and blue: 1.0 -- 1.25 keV. Intensity profiles are shown in (d) which were integrated in a RA intervall from RA = 5h49m to 5h40m. Left panel: H I component intensities, right panel: X-ray intensity in the 0.4–1.25 keV range. 
    Credit: Knies et al., A\&A, 648, 90, 2021, reproduced with permission \textcopyright ESO.
    \label{fig:Xrayspur}}
\end{figure}

\subsection{X-ray spur in the LMC}

In the LMC, a large diffuse structure called the X-ray spur 
was observed in X-rays in the \rosat\ survey performed in 1990s. 
The X-ray spur is located south of the 30 Doradus region and has an intriguing triangular shape (see Fig.\,\ref{fig:Xrayspur}). It seems to emit higher energy X-rays than what is typically expected from 
interstellar plasma in galaxies. 
\xmm\  studies of the X-ray spur have shown that the interstellar plasma 
has a higher temperature than in most other parts of the LMC, being
as high as in the region around the 
Tarantula nebula in 30 Doradus and the super-star cluster RMC 136
inside it, which are heated by a large number of very massive stars and is located north of the X-ray spur \citep{2021A&A...648A..90K}.
In the X-ray spur, however, as the complementary multiwavelength analysis shows, 
there are no indications for the past and present existence of massive stars which 
might have caused the heating of the plasma at the position. 
Instead, the X-ray spur is located between two large colliding components of 
atomic hydrogen gas in the LMC called the D (disk) and L (L-shaped) component and has most likely been caused by the collision 
of these two giant cold gas components. 
While the D-component is the main gas that fills the LMC disk, the L-component is believed to be gas stripped from the SMC through the tidal interaction with the LMC. 
The high-velocity, large-scale gas collision has most likely triggered the formation of many massive stars in the 30 Doradus region \cite{2017PASJ...69L...5F}.
The X-ray spur is well correlated with the I (intermediate)-component of the HI gas (Fig.\,\ref{fig:Xrayspur} (b)), indicating that the collision is still going on in the X-ray spur and has heated the already existing interstellar plasma by compression. This scenario is also supported by the presence of a long ridge of molecular clouds seen in CO  \cite{2008ApJS..178...56F}, starting to the south of 30 Doradus and aligning with the western end of the X-ray spur. The positions of CO clouds correlate with regions of an enhanced number of young stellar objects and suggest that also in these regions, star formation has been caused by the interaction of the large HI clouds.

\begin{figure}
    \centering
    \includegraphics[width=\textwidth,trim=0 250 0 250,clip]{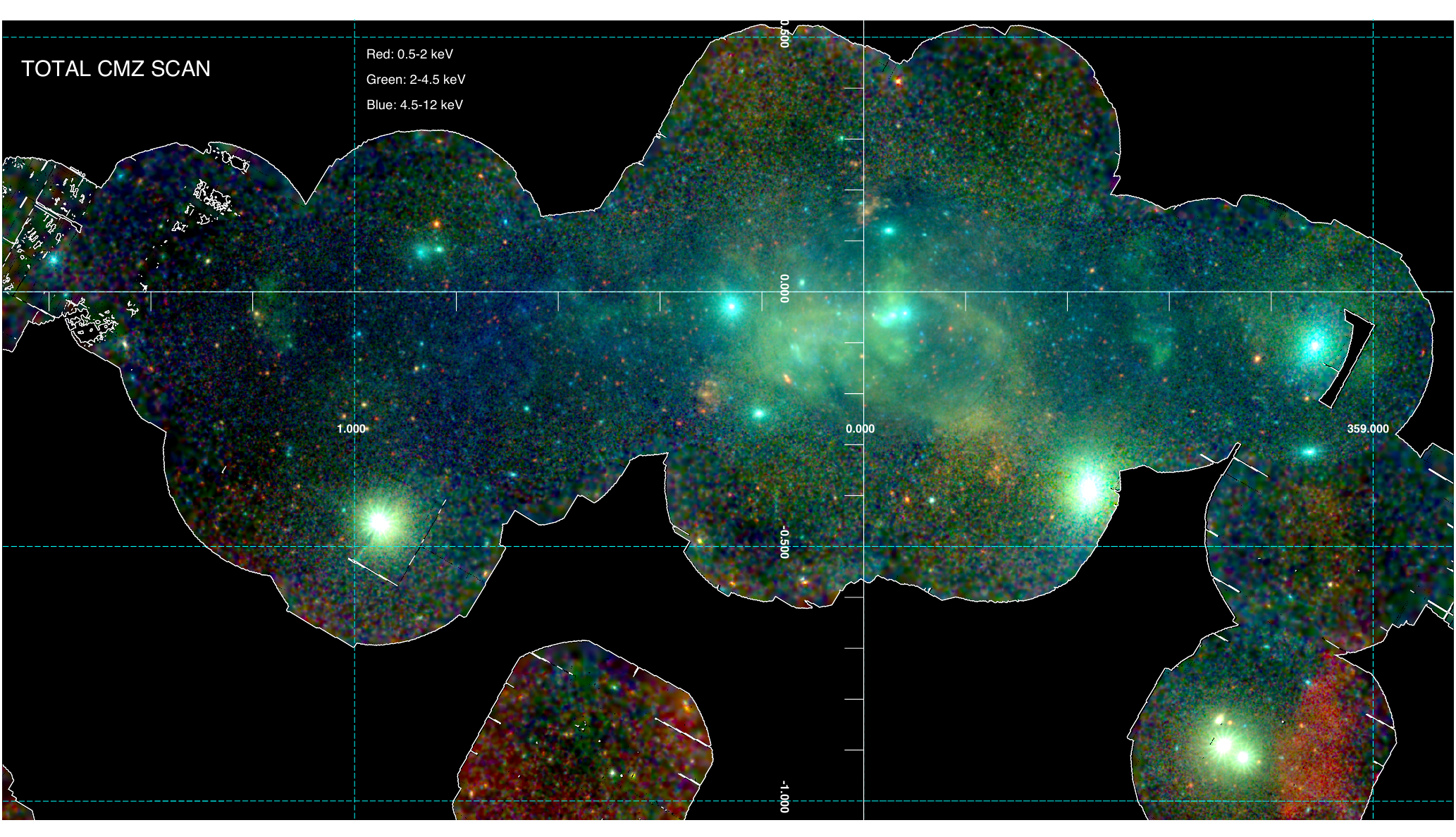}
    \caption{X-ray view of the Galactic center. In red, green and blues the emission in the 0.5-2, 2-4.5 and 4.5-12 keV energy bands is shown. This represents the mosaic image of all XMM-Newton observations within one degree of \sgras. X-ray emission from X-ray binaries, star clusters, supernova remnants, bubbles and superbubbles, HII regions, PWNs, non-thermal filaments, nearby X-ray active stars, \sgras\ and many other features are observed. 
    Adapted from \citep{Ponti15}. 
    \label{fig:GC}}
\end{figure}

\section{Galactic center}

The X-ray emission from the Galactic center region is well characterised since deep X-ray surveys of the central few degrees of the Milky Way have been performed with observatories like \xmm, \chandra, \nustar, \suzaku, among others. 

Within the central few hundred parsecs (corresponding to few degrees in extent along the Galactic plane), a peculiar environment is expected and observed. Indeed, at the center of the Milky Way there is a stellar bar with a mass of $M\sim1.9\times10^{10}$~M$_\odot$, which extends from the center to 3-5 kpc to both sides \citep{Bland-Hawthorn16} and produces a peculiar gravitational potential within $\sim0.3$ to $\sim4$~kpc from the center. 
The elongated gravitational potential makes the gas that is streaming inwards through the disc of the Milky Way reach a semi-stable elliptical orbit within the central few hundred parsecs and form the so called central molecular zone (CMZ; \citep{Morris96}). At a distance of about 8 kpc, we observe the CMZ as an extended structure  in the Galactic plane with an extent of $\sim2$ degrees. 
About $\sim3\times10^7$~M$_\odot$ of molecular gas is accumulated in the CMZ \citep{Molinari11}. 
This total mass of molecular material is high enough to induce a star-formation rate comparable to a mini-starburst region, however, star formation is rather inefficient in the CMZ.
%
\label{SFR}
Several tracers have been used to estimate the current star formation rate within the CMZ, such as supernova remnants, young stellar objects, HII regions, or young-massive stars. The star formation rate is estimated by assigning a representative age to these objects and calculating the average star formation rate over the relevant time-scale. 
These different estimates are broadly consistent with a star formation rate of $\sim0.05-0.1$ M$_\odot$ per year (see \citep{Henshaw22} for a review). 

Thousands of X-ray point sources have been detected in the CMZ \citep{Muno09}. The superior spatial resolution of \chandra\  allowed us to detect more than 9000 X-ray point sources in the direction of the CMZ \citep{Muno09}, with \sgras\ being the most famous of all such sources \citep{Baganoff03}. 
Moreover, bright diffuse X-ray emission is observed from the same region. 
Several distinct components have been identified for the diffuse X-ray emission:
X-ray reflection nebulae; X-ray filaments; the hot interstellar medium composed of supernova remnants, bubbles and superbubbles; possible remnants of tidal disruption events; an excess of hard X-ray emission claimed to be associated with the presence of a very hot plasma component; etc. \citep{Koyama18,Ponti15}. 
Diffuse emission at very high-energies (VHE; $>$100 GeV) has been detected with H.E.S.S. \citep{2018A&A...612A...9H}, which is correlated with the dense gas and clouds in CMZ.

\subsection{\sgras}

\sgras, with an estimated mass of $M_{BH}\sim4\times10^6$~M$_\odot$, is the radiative counterpart of the super-massive black hole at the center of the Milky Way \citep{Genzel10}. 
\sgras\ is often considered as the archetype of quiescent black holes. Its bolometric luminosity of $L\sim10^{35-36}$~erg~s$^{-1}$ corresponds to an Eddington ratio of $\sim10^{-8} - 10^{-9}$. 
Most of the radiation during quiescence is emitted in the sub-mm band as optically thin synchrotron radiation from a nearly thermal population of relativistic electrons with a Lorentz factor of $\gamma_e\sim10$ and electron density of $n_e\sim10^6$~cm$^{-3}$, embedded in a magnetic field with strength of $B\sim20-50$~G in a region with a size of $r\sim10$ Schwarzschild radii \citep{Genzel10}.
An even lower luminosity of $L_x\sim2\times10^{33}$ erg s$^{-1}$ is observed in the X-ray band from an extended source with a projected size of about $\sim$1'', corresponding to $\sim10^5$ Schwartzschild radii. Therefore, \sgras's extent in X-rays during quiescence is comparable to the expected size of the Bondi capture radius \citep{Baganoff03} and its X-ray emission can be reproduced by a bremsstrahlung model with an electron temperature of $kT_e\sim7\times10^7$ K and electron density of $n_e\sim100$~cm$^{-3}$ \citep{Wang13}. 
\sgras\ is located at the center of the central star cluster, which contains young and massive stars with powerful stellar winds, which can create a low-density cavity around \sgras\ itself \citep{Genzel10}. 
Simulations suggest that the X-ray emission is caused by the capture and accretion of a significant fraction of the stellar wind's material onto \sgras\ \citep{Cuadra05}.  

\sgras\ shows flares in the X-ray band with a frequency of about once per day observed as sudden increases in the X-ray luminosity by up to a couple of orders of magnitudes ($L_x\sim10^{35}$~erg~s$^{-1}$) and lasting for less than an hour to a few hours \citep{Neilsen13,Ponti15b}. The short time-scale variability implies a compact source with a size of less than a few Schwarzschild radii. Multi-wavelength campaigns have shown that synchrotron emission with a cooling break and an high-energy cut off can explain the emission during infrared to X-ray flares \citep{Ponti17,GravityColl21}. However other scenarios involving inverse Compton radiation to produce the X-ray emission have also been discussed \citep{Markoff01}.

\subsection{X-ray Reflection nebulae}

Bright X-ray emission is observed from some of the molecular clouds of the CMZ \citep{Sunyaev93,Koyama96}. A hard X-ray spectrum (with a photon index of $\Gamma\sim2$) and bright neutral Fe K$\alpha$ emission are typically observed from such sources \citep{Ponti13}.  
In theory, either cosmic-ray protons or electrons interacting with the material within the molecular clouds can cause bright X-ray emission \citep{Dogiel09}. However, the short time-scale variability observed from these sources suggests a different origin. 
As initially suggested by \citep{Sunyaev93}, the bright non-thermal X-ray emission from molecular clouds seems to be produced by the "reflection" of bright X-ray events, which occurred in the past few hundred years within the CMZ \citep{Ponti13,Churazov17}. If so, the light front of such a past event would still be propagating inside the CMZ, therefore producing reflected emission whenever it encounters a molecular cloud.
Superluminal echoes have been observed along a few molecular complexes, which support such a scenario \citep{Ponti10}. The latest observational campaigns suggest that one or more events reaching X-ray luminosities of the order of $L_X\sim10^{39}$ erg s$^{-1}$ must have occured over the past few centuries. 
Currently, only a small fraction of the clouds in the CMZ is illuminated by such a flash. 
X-ray polarimetric observations of the CMZ (with IXPE) are expected to measure the polarisation fraction of the X-ray emission from such molecular clouds \citep{Churazov17} and will allow us to verify the origin of the radiation (i.e., reflected signal) and to measure the angle between the line between the illuminating source and the cloud and the line of sight between the cloud and the observer. This geometrical information will enable us to determine the location of the illuminating source and its historical light curve, as well as to determine the three dimensional location of the illuminated clouds within the central molecular zone (see \citep{Ponti13} for a review). 


\subsection{Galactic ridge emission}

One of the largest structures with diffuse emission in the 2-10 keV band is the so-called Galactic ridge emission, which is concentrated in the Galactic disc and bulge and extends for more than 100 degrees in Galactic longitude and just a few degrees in latitude \citep{Warwick85,Revnivtsev03}. The spectrum of the Galactic ridge emission is dominated by a hard X-ray continuum plus lines (such as Fe {\rm xxv} and Fe {\rm xxvi}), which is consistent with emission from a collisionally ionised optically thin plasma with a temperature of $kT\sim 5-10$ keV \citep{Koyama07}. 
Therefore, the spectrum suggests a truly diffuse origin of this component. However, it was early realised that such a plasma would not be bound to the Milky Way and would require a source (supplied throughout the Galaxy) with a luminosity of $\sim10^{43}$ erg s$^{-1}$, which is not observed in the Milky Way \citep{Tanaka02}. 
An alternative source of the Galactic ridge emission might be a large population of faint and unresolved point sources \citep{Revnivtsev09}.

The Galactic ridge emission (traced by the Fe {\rm xxv} line and 3-20 keV continuum) is well correlated with the infrared luminosity, which traces the stellar mass density, corroborating the latter hypothesis. 
In an ultra-deep X-ray observation of a field with low extinction within $\sim1.5^\circ$ from the Milky Way center, more than 80 \% of the diffuse 6-8 keV emission has been resolved into weak point sources, such as accreting white dwarfs and coronally active stars \citep{Revnivtsev09}. The luminosity function of such objects in the Solar neighbourhood suggests that the remaining flux is consistent with being produced by even fainter sources, which, however, are below the detection threshold \citep{Sazonov06}. Therefore, it seems that the majority of the Galactic ridge emission originates from faint point sources. 

It is expected that the peak of the Galactic ridge emission is observed towards the Milky Way center due to the higher number of point sources along that line of sight. 
A peak of Fe {\rm xxv} and hard X-ray emission has indeed been observed within the central few degrees of the Milky Way. 
However, it has also been shown that the intensity of such peak appears to be about two times larger than what is expected, suggesting the presence of a truly diffuse high-temperature plasma as the origin for this excess emission \citep{Uchiyama11}. 
In deep \chandra\ observations within a few parsecs from the Galactic center about 40 \% of the emission was resolved into point sources, thus leaving the possibility open that the remaining emission is indeed produced by a very hot plasma component \citep{Yuasa12}. More data are required to clarify this.

\subsection{Hot interstellar medium}

Despite the considerably high column density of absorbing material towards the Galactic center ($N_H\sim5\times10^{22}-10^{23}$~cm$^{-2}$), which  extinguish the emission below $E\sim2-3$~keV, the emission from the hot phase of the interstellar medium can be traced through the Hydrogen and Helium-like lines of, for example, Silicon Sulfur, Argon, and Calcium \citep{Wang02,Ponti15}.  
Indeed, hot plasma with a temperature of $kT\sim0.8-1$~keV is found to pervade the Galactic center \citep{Wang02,Koyama18,Ponti15}. 
Assuming the Breitschwerdt \citep{Breitschwerdt91} potential, such plasma remains bound to the Milky Way. The radiative cooling time of the plasma within the central 50 pc is estimated to be $t_r\sim3-6\times10^6$ yr \citep{Muno04}.
The observed rate of supernovae generates enough energy to power the emission of the hot plasma. 
Additional soft X-ray emission might be produced by interaction of stellar winds from young and massive stars with the ISM \citep{Muno04} or with each other and by coronal X-ray sources. The patchy soft plasma distribution with higher concentrations (up to a factor of 10 variations within 50 pc) towards star forming regions, matches these hypotheses.


Within the central degree, two super-bubbles candidates have been observed \citep{Ponti15}. One of them (G0.1--0.1) is filled with hot plasma with an associated thermal energy in excess of $10^{51}$~erg. It is surrounded by a thick layer of cold material (the so-called Arc bubble; \citep{Simpson07}. The second candidate (G359.77--0.09) also appears to be filled with hot plasma, however its X-ray morphology is less centrally peaked and the accumulation of cold material at the edge of the super-bubble and the association with a known young star cluster is less evident. Therefore, it is not a strong candidate \citep{Ponti15}. 
The observed high plasma temperature ($kT\sim1$~keV) and small spatial extent ($r\sim15$~pc) of these super-bubble candidates suggest that they might be the youngest among a larger population of super-bubbles. Older super-bubbles have a lower plasma temperature and a lower surface brightness, and are therefore missed due to the high extinction in the soft X-ray band. 
Such a high density of super-bubbles appears rather remarkable, considered that only a few super-bubbles are know throughout the Galaxy. 


As the time-scale between consecutive tidal disruption events onto \sgras\ is expected to be a few times smaller than the time required to remove the traces of the associated energy release onto the ISM \citep{Alexander03}, it is expected that a few remnants of tidal disruption events should be observable within the CMZ. 
A few  candidates have been proposed. However, there is still no smoking gun confirming the presence of such  remnants.

\subsection{Non-thermal X-ray filaments and the Galactic center magnetic field}

Radio images of the Galactic center have revealed a large array of bright radio filaments, which extend for several tens of parsecs and a few of which have even been detected all the way up to the X-ray band \citep{LaRosa00,Heywood19,Heywood22}. 
The filaments are observed within less than $\sim1.5$ degrees from the Milky Way center and at low Galactic latitudes. They are primarily oriented perpendicular to the Galactic plane, while they appear to diverge beyond $|b|\sim0.5^\circ$ \citep{Ponti21}. 
The radio emission is consistent with Synchrotron emission from relativistic electrons. 
Since their discovery, non-thermal radio filaments have been associated with an intense, pervasive, and vertical magnetic field with a strength of $\sim1$ mG, therefore being significantly above equi-partition with the thermal energy density of the interstellar plasma.
Recent observation of arrays of filaments forming a triangular (harp-like) shape corroborate scenarios, in which  non-thermal filaments will arise if there is a source of relativistic particles that illuminates the tube of field lines in which they are trapped \citep{Thomas20}, however other models have been proposed for the origin of the non-thermal filaments.

\section{The Galactic outflow}

\begin{figure}
    \centering
    \includegraphics[width=1.1\textwidth,angle=-90]{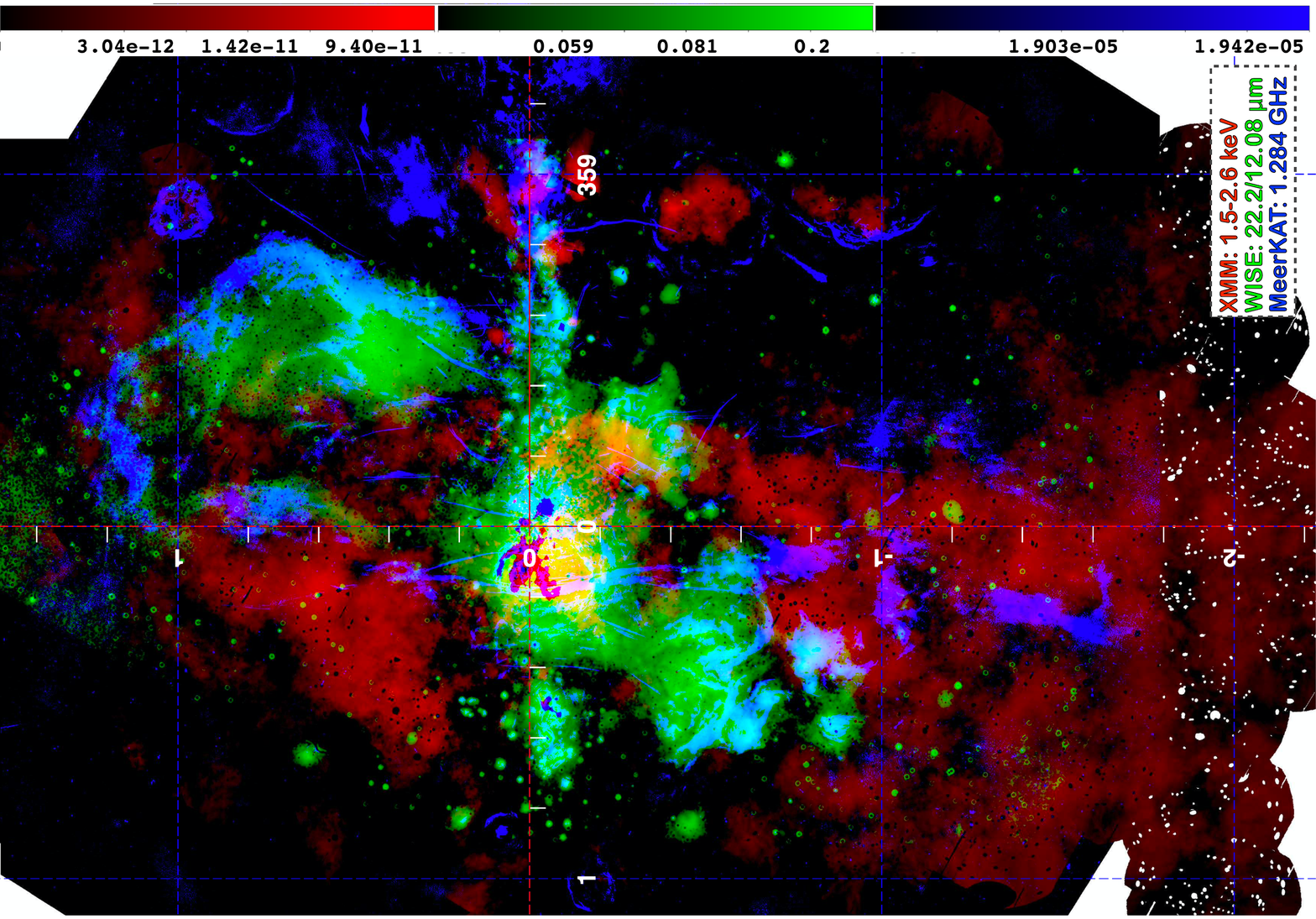}
    \caption{Multi-phase nature of the GC chimneys. Red, green, and blue show the XMM-Newton 1.5–2.6 keV, WISE (ratio of 22.2 $\mu$m/12.08 $\mu$m emission), and MeerKAT 1.284 GHz maps, respectively. This is fig.~1 from \citep{Ponti21}, Ponti, A\&A, 646, A66, 2021, reproduced with permission \textcopyright ESO.
    \label{fig:outflow}}
\end{figure}

There is a tight correlation between the properties of galaxies and their central supermassive black holes. Indeed, the feedback between the energetic activities occurring in the center of galaxies, due to the accreting supermassive black holes, and the galactic disc, due to star formation, causes large scale processes which are associated with the formation and growth of galaxies themselves. 
The consequence of such a feedback is the formation of jets and wind from the accreting supermassive black hole or winds from regions with a high specific star formation rate. The energy and momentum deposited by such winds and jets are so high that they have a profound impact on the properties of the host galaxy. Even though \sgras\ is currently in quiescence and the  star formation rate is orders of magnitudes lower than in starburst galaxies, observations have provided us with evidences for an outflow from the center of the Milky Way \citep{Bland-Hawthorn03,Su10,Ponti19,Predehl20}. 

Dynamical models of the center of the Milky Way (the bar and the CMZ) suggest a mass accretion rate of the order of $\sim1$ M$_\odot$ per year onto the central few hundred parsecs of the Milky Way \citep{Sormani19}. This number is about one order of magnitude larger than the star formation rate within the CMZ ($\sim0.1$~M$_\odot$ yr$^{-1}$; \S \ref{SFR}) and the accretion rate onto \sgras\ (which is negligibly low with $<8\times10^{-5}$~M$_\odot$ yr$^{-1}$; \citep{Genzel10}). 
The star formation rate currently observed at the Galactic center might be close to its minimum, with significantly higher rates in the past, therefore, being  far from steady state \citep{Henshaw22}.
It also implies that a large fraction of the matter that has been accreted onto the CMZ is likely expelled in a Galactic outflow. 

\subsection{Signs of a Galactic outflow}

One of the first indications for a galactic outflow was provided in observations of the motions of some HI clouds in the Galactic plane within the central $\sim10^2$ pc of the center of the Milky Way, which showed peculiar deviations from a circular motion. It was suggested that a large energy release might have occurred in the past at the Galactic center producing an outflow  \citep{Bania77,Sofue95}. 
Indeed, an event that occured a few $10^6$ years ago might have released an energy of  $E\sim10^{55}$ erg causing the observed kinematics of the so-called expanding molecular ring \citep{Sofue95} and enough power to push away $\sim10^7$ M$_\odot$ of material. 
However, soon it was demonstrated that non-circular motions are expected as a consequence of the stellar bar at the center of the Milky Way \citep{Binney91}. Indeed, the bar dominates the gravitational potential within $\sim10^2$ pc from the center and makes the gas  move along on very elongated orbits, which significantly deviate from circular motions \citep{Contopoulos89}.


Detailed analysis of the \rosat\ and the Midcourse Space Experiment maps of the Galactic center \citep{Bland-Hawthorn03} has revived the discussion of a Galactic outflow, presenting some evidence of mass ejections on scales of degrees (several $10^2$~pc) to tens of degrees (several kiloparsecs). The infrared map shows spurs emerging from the Galactic plane on scales of several $10^2$ pc indicating the presence of an outflow. Dust entrained in a large-scale, bipolar wind would produce infrared spurs, while the same outflow would be the origin of the faint edge brightened X-ray emission seen in the \rosat\ maps on kiloparsecs scales \citep{Bland-Hawthorn03}.

A few years later, the discovery of the \fermi\ bubbles made the Galactic outflow scenario even more compelling \citep{Su10}. 
The \fermi\ bubbles are a pair of large scale structures, centered on the center of the Milky Way and extending up to 50 degrees (i.e. $\sim10$ kpc) above and below the Galactic plane. They are detected clearly in the hard gamma-ray band ($E>2$~GeV) with a uniform gamma-ray surface brightness with sharp edges with a luminosity of $\sim4\times10^{37}$ erg s$^{-1}$ and an estimated energy content of $E\sim10^{55}$ erg \citep{Su10}. This gamma-ray emission is believed to be produced by the cosmic rays associated with the Galactic outflow. 

Additionally, an excess of radio emission, the so-called Galactic haze has been found in the {\it wmap} and the {\it planck} data and seems to be associated to the outflow \citep{Dobler10}. The hard microwave emission is consistent with a power law emission with a spectral index of $2.55\pm0.05$. This spectrum is inconsistent with free-free emission and favors synchrotron radiation from a power law electron population. 
The microwave haze appears to be connected to the gamma-ray features and strongly suggests that  these cosmic rays are also tracers of the Galactic outflow. 

In addition, large scale maps of the S-PASS survey have  revealed two giant, linearly polarized radio lobes, containing three ridge-like substructures, which appear to be connected to the X-ray and gamma-ray features. Therefore, these radio lobes also seem to be connected to the Galactic outflow \citep{Carretti13}. These features are permeated by a strong magnetic field of up to 15 $\mu$G. They have been interpreted as the result of an outflow which carries a large amount of magnetic energy into the Galactic halo, corresponding to $\sim10^{55}$ ergs, similar to the energy of the outflows traced at other wavelengths. 


The UV spectra of stars and quasars in the background and foreground of the \fermi\ bubbles have revealed absorption features, which are characteristic of warm plasma and are not observed in the spectra of quasars outside the bubbles \citep{Fox15,Bordoloi17}.
The plasma has most likely been transported with the Galactic outflow.
The excellent resolution of UV spectra makes a measurement of the kinematics of the gas possible. The observed kinematics appear to be consistent with a global outflow of the warm plasma with velocities of the order of $\sim10^3$~km~s$^{-1}$. If so, the Galactic wind creating the \fermi\ bubbles must have started $\sim6-9$~Myr ago \citep{Bordoloi17}. 
Recent observations of the Wisconsin H-Alpha Mapper have revealed emission from ionised hydrogen and nitrogen at similar speeds, which can be associated with the same outflow \citep{Krishnarao20}. 


Recently, an array of about 200 cool gas clouds have also been discovered in HI observations at the base of the \fermi\ bubbles \citep{DiTeodoro18}. The clouds seem to move at speeds of few $10^2$~km~s$^{-1}$. The further away they are located from the center, the faster they seem to become. Detailed observations of two of such HI clouds have revealed the presence of molecular gas in their cores \citep{DiTeodoro20}, which is surprising if the current activity of \sgras\ or star formation powers the outflow.  
The observations of warm, cold, and molecular clouds of hydrogen and nitrogen emission in addition to the flux in the $\gamma$-ray band (from the cosmic rays) show that the Galactic outflow has a truly multi-phase nature. 

\subsection{The Chimneys and the base of the Galactic outflow}

The edges of the Galactic outflow are well visible from high Galactic latitudes down to few degrees from the Galactic plane in $\gamma$-ray, X-ray, and infrared maps. By extrapolating these edges it appears that the Galactic outflow originates from the CMZ, and hence within the central few hundred parsecs of the Galactic center. 
Previous studies, however, yielded no indications for the Galactic outflow being connected to the Galactic center \citep{Bland-Hawthorn03,Su10}. 
Deep X-ray observations of the CMZ extending to $\sim1-2$ degrees in latitude have recently revealed the presence of two chimneys of hot plasma centered on \sgras\  with a base in the Galactic plane with an extent of of few tens of parsecs and extending vertically for more than few hundred parsecs \citep{Ponti19,Ponti21}. High-sensitivity MeerKAT observations of the same region have revealed two edge-brightened lobes, which seem to surround the X-ray emission \citep{Heywood19}. 
The X-ray and radio features are most likely strongly connected, being two faces of the same process \citep{Ponti21}. 
\suzaku\ observations of the emission south of the Galactic center revealed that the emission is dominated by that of recombining plasma, which is most likely a relic of past actitivies in the Galactic center region
\citep{2013ApJ...773...20N}.
While it seems clear that the chimneys are powered by an outflow from the Galactic center, the source of such an outflow (and whether some of the X-ray and radio structures are associated with foreground features) is still unclear. 
In particular, the rapid drop of the pressure of the hot plasma, which is  observed to occur from few to few tens of parsecs from \sgras, implies that there is a rapidly expanding outflow within the so-called Sgr A's bipolar lobes \citep{Morris03,Ponti15,Ponti19}. 
However, the shallow pressure profile observed within the chimneys suggests that the chimneys are the product of an outflow which is either still going on at low power or has already left the central hundred of parsecs. The X-ray and radio emissions are probably only the vestige of the passage of a much more powerful outflow which occurred in the past \citep{Ponti19,Ponti21}. 
In any case it is clear that the chimneys represent the channel which connects the activity in the central parsecs with the base of the Galactic outflow. 

\subsection{The \erosita\ bubbles}

The first all-sky X-ray map of the \erosita\ telescope has made it evident that the north polar spur is the brightest part of a large bubble in the northern Galactic sky, which is emerging from the Galactic center. 
The high-sensitive X-ray map has additionally revealed an almost perfectly symmetric southern counterpart. Together, both bubbles have been named  \erosita\ bubbles and they seem to extend about 15~kpc above and below the Galactic center. Being significantly larger and more energetic than the \fermi\ bubbles \citep{Predehl20}, the \erosita\ bubbles possess a volume and energy 
about ten times larger than the \fermi\ bubbles (however this estimate depends on the assumed geometry of the bubbles). The roughly constant X-ray surface brightness of the \erosita\ bubbles is consistent with a thick shell morphology, possibly associated with shock-heated plasma at the edge of the Galactic outflow. 
Despite the double bubble morphology centered at the Galactic center and the significant X-ray absorption over most of the \erosita\ bubbles, which strongly suggest that they originate from the Galactic center, it has been argued that some part of these structures might be local, within less than a kiloparsec from us \citep{Lallement22}. Indeed, a super-position of several different features is very likely for such an extended X-ray emission that covers a good fraction of the entire sky. 


Recently, it was reported that the extent of the \erosita\ bubbles along the Galactic disc is comparable to that of a cavity in the HI disc \citep{Sofue21}. The vertical extent of the HI disc is significantly reduced within the central $\sim3$~kpc \citep{Lockman84,Lockman16}. It is possible that the formation of such a crater in the HI disc has been formed by the shock wave from the Galactic center outflow, which have swept away the upper layer of the gas in the Galactic disc \citep{Sofue21}, although alternative hypotheses also remain. 


Since the discovery of the Galactic outflow, numerous physical constraints have been accumulated, allowing us to get a deeper understanding of the phenomenon. 
The sum of observational constraints indicate that the Galactic outflow is either the result of a past AGN-like activity of \sgras\ or that of a burst of star formation, with a rate larger than what is currently observed  \citep{Cheng11,Crocker11,Crocker15,Kataoka18}. 
Indeed, an outflow can be generated in both scenarios.  

\section{Summary}

In this chapter, we have presented results from observations of the hot phase of the ISM, which are mainly carried out in X-rays. The findings are often corroborated by observations at other wavelength bands, and theoretical considerations and numerical simulations.

The main cause for the hot phase of the ISM are stellar winds of massive stars and in stellar clusters and supernova explosions. In addition, wind-wind collisions in stellar clusters are also believed to be efficient in heating of the ISM. Soft X-ray emission from thermal plasma is detected in SNRs and inside HII regions and superbubbles in the Milky Way and the nearby galaxies. Emission from non-thermal particles accelerated in the interstellar shocks is also expected to be observed not only in radio, but also at higher energies. Emission from non-thermal electrons has been observed and confirmed at X-ray to gamma-ray energies in the superbubble 30 Doradus C in the Large Magellanic Cloud.
In the future, more sensitive observations will become possible with CTA.
X-ray observations towards HII regions and superbubbles also allow to measure the column density of cold ISM on the line of sight due to the absorption of emission from the hot interstellar plasma.
Recent multi-wavelength observations of the ISM in the Milky Way and nearby galaxies have also revealed that collisions of large interstellar clouds can cause the heating of interstellar plasma and the formation of massive star clusters.

We have also in particular discussed the Galactic center region, in which X-ray surveys have been carried out to study the sources and the ISM around the nuclear supermassive black hole Sir A*. In these studies, various interstellar structures with X-ray emission have been found: filaments, reflection nebulae, SNRs, bubbles, and superbubbles. The X-ray emission from Sgr A* is dominated by synchrotron emission from nearly thermal electrons accompanied with X-ray flares on timescales of days. In the Galactic plane, the Galactic ridge emission has been detected with emission above 2 keV, which would correspond to plasma temperatures higher than what is typically expected in interstellar plasma. Dedicated X-ray studies have shown that the emission is to a large extent caused by unresolved point sources like accreting white dwarfs and coronally active stars. Outflows driven by the supermassive black hole and/or massive stars due to enhanced star formation are also found from the nuclear region. In addition to observations of structures in X-rays and dust emission, the detection of the Fermi bubbles at GeV energies provided us with a clear indication of such outflows, also corroborated with observations in radio, sub-mm, Infrared, UV. Recently the detection of the eROSITA bubbles in X-rays confirmed the existence of hot plasma in the Galactic halo caused by outflows from the Galactic center region.

\section{Acknowledgments}

MS acknowledges support from the Deutsche Forschungsgemeinschaft through the grants SA 2131/13-1, SA 2131/14-1, and SA 2131/15-1.
GP acknowledges funding from the European Research Council (ERC) under the European Union’s Horizon 2020 research and innovation programme (grant agreement No 865637). 
JM acknowledges support from a Royal Society-Science Foundation Ireland \emph{University Research Fellowship} (20/RS-URF-R/3712) and an Irish Research Council \emph{Starting Laureate Award} (IRCLA\textbackslash 2017\textbackslash 83).


\bibliographystyle{apj} 
\bibliography{mackey,ponti,sasaki}

\end{document}